\documentclass[aps,prl,twocolumn,floatfix]{revtex4-2}
\usepackage{amssymb}
\usepackage{physics}
\usepackage{graphicx}
\usepackage{times}
\usepackage{amsmath}
\usepackage{amsthm}
\usepackage{amsfonts}
\usepackage[T1]{fontenc}
\usepackage[latin9]{inputenc}
\usepackage{array}
\usepackage{multirow}
\usepackage{color}
\usepackage{esint}
\usepackage{bm}
\usepackage{color}
\usepackage{bbm}
\usepackage{hyperref}
\usepackage{url}
\usepackage{babel}

\newcommand{\bsub}{\begin{subequations}}
\newcommand{\esub}{\end{subequations}}

\usepackage{float}
\hypersetup
{	colorlinks,%
	citecolor=green,%
	linkcolor=red,%
	urlcolor=blue%
}
\allowdisplaybreaks[4]
\begin{document}
\title{Stripe-tuned superconductivity in single-flavor metals with nontrivial quantum geometry}
\author{Yi-Ting Tu}
\affiliation{Condensed Matter Theory Center and Joint Quantum Institute, Department of Physics, University of Maryland, College Park, Maryland 20742, USA}
\author{Yang-Zhi Chou}
\affiliation{Condensed Matter Theory Center and Joint Quantum Institute, Department of Physics, University of Maryland, College Park, Maryland 20742, USA}
\author{Yi Huang}
\altaffiliation{Present address: Department of Materials Science and Engineering,
University of Washington, Seattle, Washington 98195, USA}
\affiliation{Condensed Matter Theory Center and Joint Quantum Institute,
Department of Physics, University of Maryland, College Park,
Maryland 20742, USA}
\author{Sankar Das Sarma}
\affiliation{Condensed Matter Theory Center and Joint Quantum Institute, Department of Physics, University of Maryland, College Park, Maryland 20742, USA}

\begin{abstract}
We study how the interplay between nontrivial quantum geometry and an applied stripe potential affects superconductivity in a two-dimensional single-flavor metal. Assuming a weak contact attractive interaction and focusing on the lowest subband in the presence of a strong stripe potential, we analytically derive two possible pairing states in the quasi-one-dimensional limit. In addition to the conventional longitudinal $p_y$-wave order (with the stripes along the $y$ direction), we find that an exotic transverse $p_x$-wave order can be stabilized. The competition between these two orders is controlled by the electron density of each stripe and the Berry-curvature-dressed interaction. Notably, the transverse $p_x$ wave order develops a nodal line at $k_x=0$, while the longitudinal $p_y$ order is fully gapped. We discuss the possible experimental probes distinguishing these orders. Our results establish a way of controlling the pairing symmetry through a stripe potential, predicting superconductivity with nontrivial quantum geometry.
\end{abstract}

\maketitle

\begin{figure}[t]
    \includegraphics[scale=0.3]{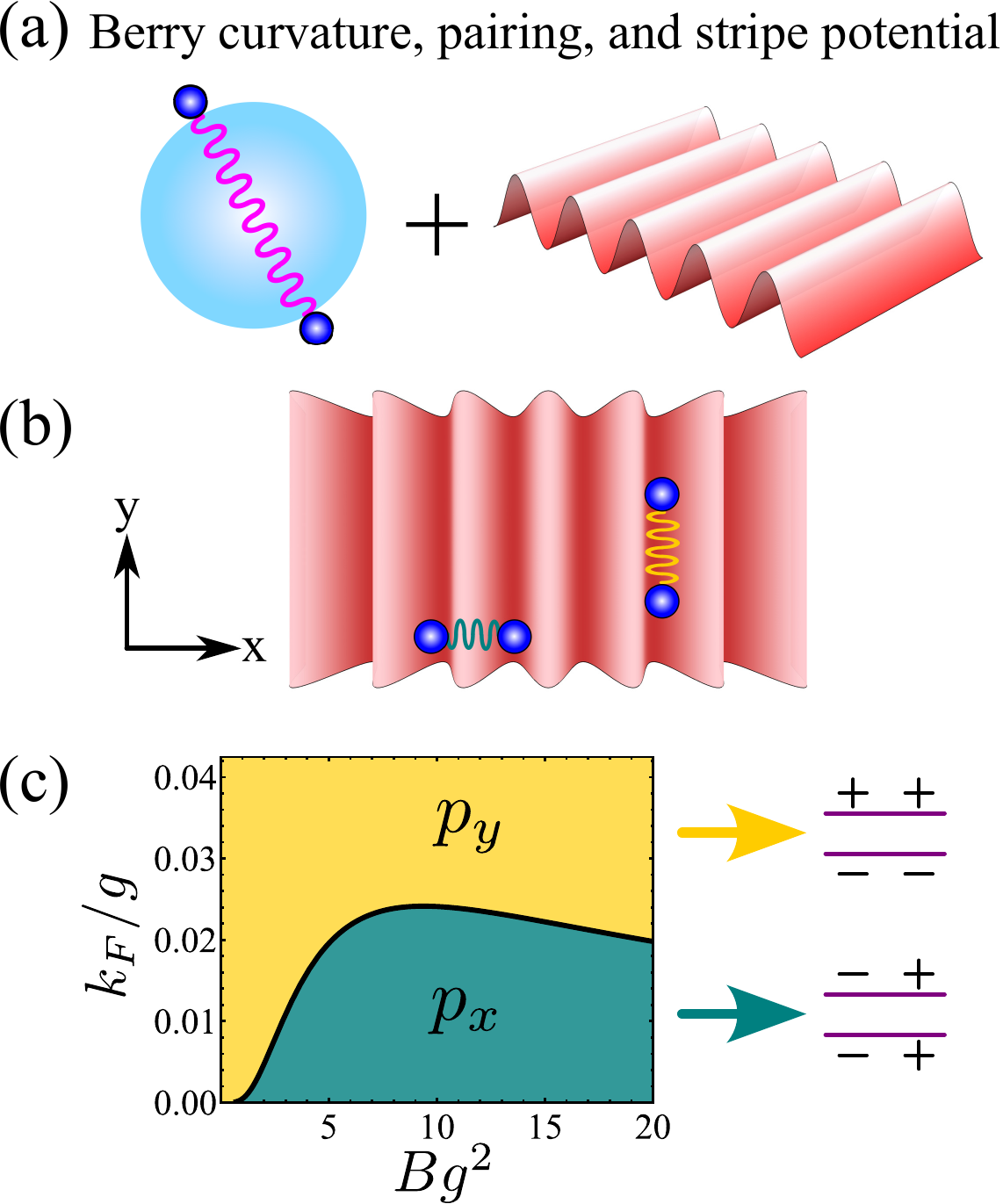}
    \caption{
    Summery of results. (a) Setup: We consider a 2D single-flavor electron systems with a finite Berry cuarvature, an attractive interaction, and a stripe potential. (b) The real-space top view of the system and two different types of effective pairings. Each stripe forms a 1D subsystem along the $y$ direction. The intra-stripe pairing (marked by the yellow wiggly line) is caused by the $q_y$ dependence of the effective interaction $V_\mathbf{q}$; the inter-stripe pairing (marked by the teal wiggly line) is caused by the $q_x$ dependence of the effective interaction $V_\mathbf{q}$. (c) Asymptotic phase diagram in the weakly interacting and strong stripe limit. The yellow and teal regions correspond to $p_x$ and $p_y$-wave pairing (with sign of pairing function on the two-line FS shown next to the phase digram). Strictly speaking, two regions are separated by a sharp crossover, but the sharp crossover can be treated as a phase transition in the strong stripe limit. All plots use a fix effective stripe strength $Ag^2=0.2$.} 
    \label{fig:intro}
\end{figure}

\textit{Introduction. ---} Engineering band structures through superlattice potentials is a promising strategy for exploring novel quantum phenomena in modern condensed matter physics~\cite{Andrei2021,Sun2024,Andrei2020,Balents2020,Mak2022,Li2026}. Most existing studies focus on creating moir\'e systems by stacking multiple two-dimensional layers with angular mismatch (e.g., magic-angle twisted bilayer graphene~\cite{Cao2018,Cao2018a} and twisted WSe$_2$~\cite{Wang2020}) or lattice mismatch (e.g., AB stacked WSe$_2$/WS$_2$~\cite{Tang2020,Regan2020}). Additionally, imposing external periodic potentials (e.g., by placing a patterned dielectric~\cite{Sun2024a} or aligning with a substrate~\cite{Dean2013,Ponomarenko2013}) on existing materials can also induce moir\'e potentials, thereby reconstructing the electronic structure and correlated states~\cite{Cano2021,Chou2021,Ghorashi2023,Ghorashi2023a,Zeng2024,Sun2024a}. One prominent example is the fractional quantum anomalous Hall effect in rhombohedral multilayer graphene aligned with hBN~\cite{Lu2024}, where substrate alignment generates a moir\'e superlattice that reconstructs the active band.

Recently, a quarter-metal superconductivity (SC) has been reported in rhombohedral tetralayer and pentalayer graphene systems \cite{HanT2025a}. Since the normal state is spin- and valley-polarized (i.e., having only a single fermion species), $s$-wave pairing is forbidden, making it a candidate for a chiral $p$-wave SC~\cite{Read2000,Alicea2012,Kallin2016} due to the finite Berry curvature in the active band \cite{GeierM2024,YangH2025a,May-MannJ2025}. This possibility has motivated a number of theoretical studies \cite{ChouYZ2025,GeierM2024,YangH2025a,May-MannJ2025,Jahin2026,Patri2025,Li2025,Wang2026,Chou2025,Chen2025,Yoon2026,Qin2026,DongZ2025,GilA2025,Zhu2026,Sau2024,Parra-MartinezG2025a,Christos2025}, trying to understand the origins of pairing and other basic properties. A subsequent experiment \cite{MorissetteE2025b} found a quarter-metal SC with strong transport anisotropy, suggesting a scenario in which SC coexists with a stripe potential (either intrinsically or externally induced). Regardless of the origin of the stripe potential, this new development leads to an interesting and general question: How does a stripe superlattice potential affect pairings of electrons in the presence of finite Berry curvatures. 

In this Letter, we develop an analytically solvable model to study how the interplay between nontrivial quantum geometry and stripe potential affects SC in a single-flavor metal.
We consider a $k^2$-dispersing band with uniform Berry curvature~\cite{TanT2024a,Wang2026,May-MannJ2025,Li2025,Jahin2026,Bernevig2025,Chou2025}, a preexisting real-space stripe potential, and a contact attractive interaction. The setup is illustrated in Fig.~\ref{fig:intro}(a).
By projecting the interaction onto the lowest, 1D-like subband due to the stripe and solving for the linearized gap equation (LGE), we find that the most stable pairing can be either transverse $p_x$-wave-like or longitudinal $p_y$-wave-like, where the stripes are along the $y$ direction, depending on the Berry curvature $B$ and Fermi momentum $k_F$ [Fig.~\ref{fig:intro}(c)]. 
The results can be understood as the competition between the inter-stripe and intra-stripe pairings [illustrated in Fig.~\ref{fig:intro}(b)], which arise naturally in our model with Berry curvature.
We also numerically calculate the weak-stripe regime beyond the analytically solvable limit, showing that the boundary between the $p_x$- and $p_y$-like pairing blurs out with a crossover towards the $p+ip$ chiral SC in the absence of stripes.
Our work provides an unprecedented way to manipulate the pairing symmetry of SC through the external stripe potential.

\textit{Model. ---} 
We start with a model of a parent electronic band containing only a single fermion species, with a constant Berry curvature $B>0$ and the ideal form factor~\cite{TanT2024a,Wang2026,May-MannJ2025,Li2025,Jahin2026,Bernevig2025,Chou2025}
\begin{equation}
    F(\mathbf{k}',\mathbf{k})=e^{-\frac{B}{4}\left(|\mathbf{k}'-\mathbf{k}|^2+2i\mathbf{k}'\times\mathbf{k}\right)},
\end{equation}
where $\mathbf{k}'\times\mathbf{k}=k'_xk_y-k'_yk_x$.
For analytical tractability, the band structure is assumed to be parabolic: $\mathcal{E}(\mathbf{k})=k^2/2m$.
For the stripe order, we assume a pre-existing potential $U(x,y)=-2U_0\cos gx$ in real space, corresponding to $U_{\mathbf{g}}=-U_0$ if $\mathbf{g}=(\pm g,0)$, and $0$ otherwise, in momentum space.
The non-interacting Hamiltonian is
\begin{equation}
    H_0=\sum_{\mathbf{k}}c^\dagger_\mathbf{k}\mathcal{E}(\mathbf{k})c_\mathbf{k}+\sum_{\mathbf{k},\mathbf{k}'}c^\dagger_{\mathbf{k}'}U_{\mathbf{k}'-\mathbf{k}}F(\mathbf{k}',\mathbf{k})c_\mathbf{k},
\end{equation}
which is a tight-binding model in $k$ space with the hopping strength given by $\exp(-\frac{1}{4}Bg^2)U_0$. The most obvious effect of the Berry curvature here is to suppress the bare stripe potential $U_0$.
We define the effective strength of the stripe and a gauge transformation
\begin{equation}
    \tilde U_0=e^{-\frac{B}{4}g^2}U_0,\quad \tilde c_\mathbf{k}=e^{i\frac{Bk_xk_y}{2}}c_\mathbf{k}
\end{equation}
so that the $k$-space tight binding model simplifies to
\begin{equation}\label{eq:H0transformed}
    H_0 = \sum_{\mathbf{k}}\frac{k^2}{2m}\tilde c^\dagger_\mathbf{k}\tilde c_\mathbf{k} - \tilde U_0\sum_\mathbf{k}\left(\tilde c^\dagger_{\mathbf{k}+(g,0)}\tilde c_\mathbf{k}+\tilde c^\dagger_{\mathbf{k}-(g,0)}\tilde c_\mathbf{k}\right).
\end{equation}
Note that the ``potential'' of this $k$-space model is $k^2/2m$. In the limit $g\to 0$, the model can be approximated as a harmonic oscillator, whose ground-state wavefunction has the Gaussian form $|\psi(k_x)|^2\sim\exp(-2Ak_x^2)$, where 
\begin{equation}
    A=\frac{1}{2g\sqrt{2m\tilde U_0}}
\end{equation}
is the parameter that characterizes the (inverse) effective strength of the stripe potential.

\begin{figure}
    \centering
    \includegraphics[trim={10 0 5 0},width=\linewidth]{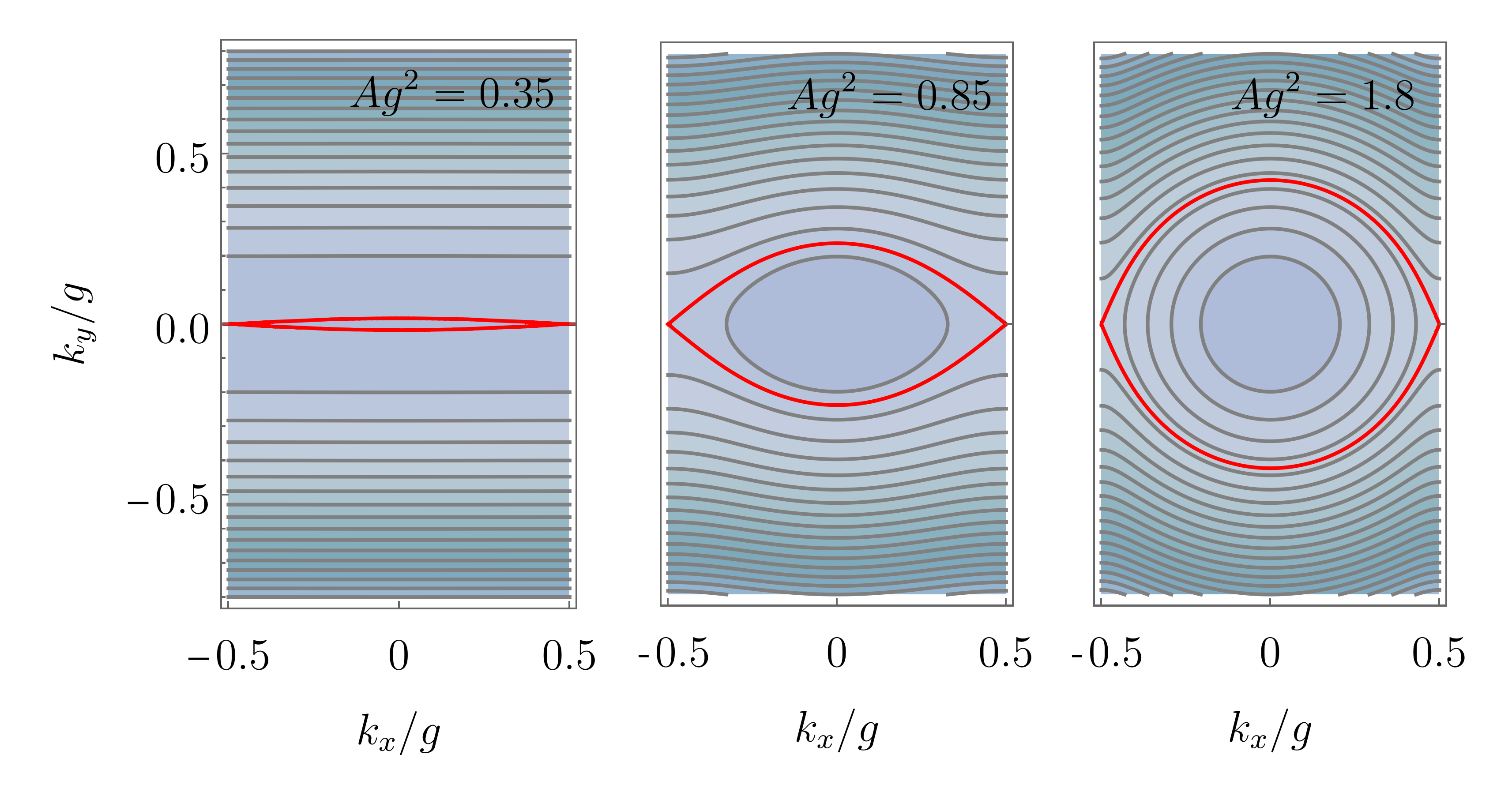}
    \caption{The structure of the lowest subband. Gray countours indicates $m(E-E_0)/g^2=0.02,0.04,\ldots$, and red countours indicates the one passing the VHS.}
    \label{fig:band}
\end{figure}

\textit{Lowest subband. ---}
Since our Hamiltonian is periodic in the $x$ direction, the parent band (which is assumed to be continuous in both directions) is folded to have a Brillouin zone of size $g$ in the $x$ direction and is still infinite in the $y$ direction.
The spectrum of the lowest subband is plotted with three representative values of $A$ in Fig~\ref{fig:band}.
In the $A\to0$ limit, the stripe potential makes the system effectively 1D-like, so that the FS approaches two horizontal straight lines.
On the other hand, when $A\to\infty$, the stripe disappears, so that the dispersion approaches that of the 2D parent band with a single circular FS.
In this work, $k_F$ is defined by the value on the $+k_y$ axis at its intersection with the FS.
Thus, for a given $A$, we have an open FS consisting of two curves for large $k_F$, and a closed FS for small $k_F$, separated by a critical $k_F$ at which the FS passes through the Van Hove singularity (VHS) at $\mathbf{k}=(g/2,0)$.
Note that for larger $A$, the Fermi surface can involve higher subbands. To simplify the problem, we will restrict ourselves to $k_F<g/2$ so that only the lowest subband is needed.
Generalization to more subbands is straightforward, but tedious.

For $Ag^2\ll 1$, the system in real space can be treated as a tight-binding model in the $x$ direction, allowing for analytical descriptions of the energy and wavefunctions.  (See Supplemental for the general case.)
Due to the strong stripe potential, the cosine variation of the energy along the $x$ direction is negligible, leading to a dispersion only in the $y$-direction.
\begin{equation}\label{eq:real-space-TB}
    E\approx E_0+\frac{k_y^2}{2m}
\end{equation}
where $E_0$ is the ground state energy,
and the FS can be approximated as two straight lines $k_y=\pm k_F$.
Moreover, the wavefunction can be approximated by a Gaussian
\begin{equation}\label{eq:wavefunction}
    \hat\psi_{n,\mathbf{k}}=\left(\frac{2Ag^2}{\pi}\right)^{\frac{1}{4}}e^{-A(k_x+ng)^2-i\frac{B}{2}(k_x+ng)k_y},
\end{equation}
where $\mathbf{k}$ is in the first BZ ($k_x\in[-g/2,g/2)$, $k_y\in\mathbb{R}$) and $n\in\mathbb{Z}$ is the internal index due to band folding.
This $B$-dependent phase is the Berry-phase-induced effect other than stripe potential suppression.

Our model is characterized by three dimensionless parameters: $Ag^2$, $Bg^2$, and $k_F/g$.
Note that the suppression of effective stripe strength due to the Berry curvature is absorbed into $Ag^2$, so $Bg^2$ characterizes the effect of $B$ that produces the complex phase in the single-particle wavefunctions.

A similar construction exists in Ref.~\cite{TanT2024a}, where they also considered an ideal parent band with constant Berry curvature and a real space periodic potential, and constructed the subbands using a $k$-space tight binding model.
However, their potential is periodic in both directions, and they focus on the quantum geometry of the subbands rather than SC, which is our focus.

\textit{Projected interaction. ---}
We consider a BCS-type point-like attractive interaction in the parent band
\begin{multline}\label{eq:Hintq}
    H_\text{int}=-\frac{V_0}{2\mathcal{A}}\sum_{\mathbf{k}_1,\mathbf{k}_2,\mathbf{q}}F(\mathbf{k}_1,\mathbf{k}_1+\mathbf{q})F(\mathbf{k}_2,\mathbf{k}_2-\mathbf{q})\\\times c^\dagger_{\mathbf{k}_1}c^\dagger_{\mathbf{k}_2}c_{\mathbf{k}_2-\mathbf{q}}c_{\mathbf{k}_1+\mathbf{q}}
\end{multline}
where $V_0$ is the interaction strength and $\mathcal{A}$ is the 2D area of the system.
This type of interaction may be due to phonons or other pairing glues, but here we only introduce it to study the effect of stripe and Berry curvature on pairing.
Also note that although a static contact interaction does not exist in a single-component electron gas, it can lead to pairing either through the nontrivial quantum geometry (as in our case)~\cite{May-MannJ2025,Patri2025,Li2025,Wang2026} or dynamical effects~\cite{Chou2025}.

By projecting $H_\text{int}$ onto the lowest subband, we obtain the effective interaction under the strong-stripe approximation. See Supplemental Material (SM) \cite{SM} for a derivation.
\begin{multline}\label{eq:VFinal}
    V_{\mathbf{k}_1,\mathbf{k}_2,\mathbf{q}}=V_0\sum_m\exp\left\{-\left(A+\frac{B}{2}\right)(q_x+mg)^2\right\}\\
    \sum_N\exp\Bigg\{i\frac{2\pi N}{g}(k_{1x}-k_{2x}+q_x)\\
    -\frac{1}{4Ag^2}(Bgq_y+2\pi N)^2-\frac{B}{2}q_y^2\Bigg\}.
\end{multline}

To interpret this effective interaction, note that the two sums are independent, and that the interaction depends on the momentum transfer $\mathbf{q}$ and $k_{1x}-k_{2x}$ (the phase factor in the second line).
So roughly, the interaction is a sum of Gaussian peaks at a grid $\mathbf{q}\approx(mg,N\frac{2\pi}{Bg})$ in the momentum-transfer space.
The $q_x$ direction can be understood as summing the peaks from different BZs, and if $B$ is not too large, it reduces to a slight variation across the BZ, as the width of the Gaussian is comparable to or much larger than the size of the BZ.
For the $q_y$ direction, we can treat the $\exp(-\frac{B}{2}q_y^2)$ as an envelope for the peaks approximately at integer multiples of $\frac{2\pi}{Bg}$.
If $B$ is not too large, the interaction is dominated by the $N=0$ peak, with all the finite-$q_y$ peaks suppressed by the Gaussian envelope.
Note that the $k_{1x}-k_{2x}$ part occurs in the $N$-dependent phase shifts along $q_x$ in the finite-$q_y$ peaks. (See SM for an extended discussion with full plots \cite{SM}.)

If we approximate the variation along $q_x$ up to the first harmonic and only keep the central peak ($N=0$) along $q_y$, the effective interaction can be simplified as
\begin{align}\label{eq:VN0}
    V_{\mathbf{q}}=V_0 c_1\left(1+2c_2\cos\frac{2\pi q_x}{g}\right)c_3(q_y)
\end{align}
where 
\begin{subequations}\label{eq:c_s}
   \begin{align}
    c_1=&\sqrt{\frac{\pi}{\left(A+\frac{B}{2}\right)g^2}},\\
    c_2=&e^{-\frac{\pi^2}{\left(A+\frac{B}{2}\right)g^2}},\\ 
    c_3(q_y)=&e^{-\left(\frac{B^2}{4A}+\frac{B}{2}\right)q_y^2}.
\end{align} 
\end{subequations}
Note that the interaction now depends only on the momentum transfer $\mathbf{q}$ but not on the individual momenta. Equation~(\ref{eq:VN0}) is anisotropic along the $x$ and $y$ directions. The coupling along the $x$ direction is only between the nearest-neighbor stripes (i.e., the Fourier transform of the $\cos(2\pi q_x/g)$ term), while the coupling along the $y$ direction is Gaussian-like. As a result, tuning the Berry curvature $B$ has different effects on the couplings along the two directions. Moreover, $V_\text{q}$ is real and positive.
This approximated interaction will be used for the analytical pairing calculations below.

\textit{Pairing. ---}
Now, we study the possible pairings driven by the effective interaction $V_{\mathbf{q}}=V_{\mathbf{k}'-\mathbf{k}}$ [Eq.~(\ref{eq:VN0})] in the quasi-1D limit, focusing on $B>0$.
Although the pairing depends on the interaction in the entire $\mathbf{k}$-space, as a first approximation, we can assume that $T_c$ is small (corresponding to small $V_0$), so that pairing is dominated by the interaction with $\mathbf{k},\mathbf{k}'$ on the FS. In SM \cite{SM}, we solve the problem numerically and incorporate contributions away from the FS. The qualitative results remain the same. In the following, we focus on the analytical results within the FS-only approximation.

To have an analytical solution in the quasi-1D limit, we approximate the FS as two straight lines $k_y=\pm k_F$. 
Writing $V_{\mathbf{k}'-\mathbf{k}}$ as $V_{\mathbf{k},\mathbf{k}'}$ and $V_{(k_x,sk_F),(k'_x,s'k_F)}$ as $V^{s,s'}_{k,k'}$ (where $s,s'$ are $\pm$ signs), the interaction becomes
\begin{align}
    V^{++}_{k_x,k'_x}&=V^{--}_{k_x,k'_x}= V_0  c_1\left(1+2c_2\cos\frac{2\pi (k_x'-k_x)}{g}\right)\label{eq:Vqcut1}\\
    V^{+-}_{k_x,k'_x}&= V^{-+}_{k_x,k'_x}=V^{ss}_{k_x,k'_x} c_3(2k_F).\label{eq:Vqcut2}
\end{align}

\begin{figure}
    \centering
    \includegraphics[width=0.75\linewidth,trim=0 0.4cm 0 0cm]{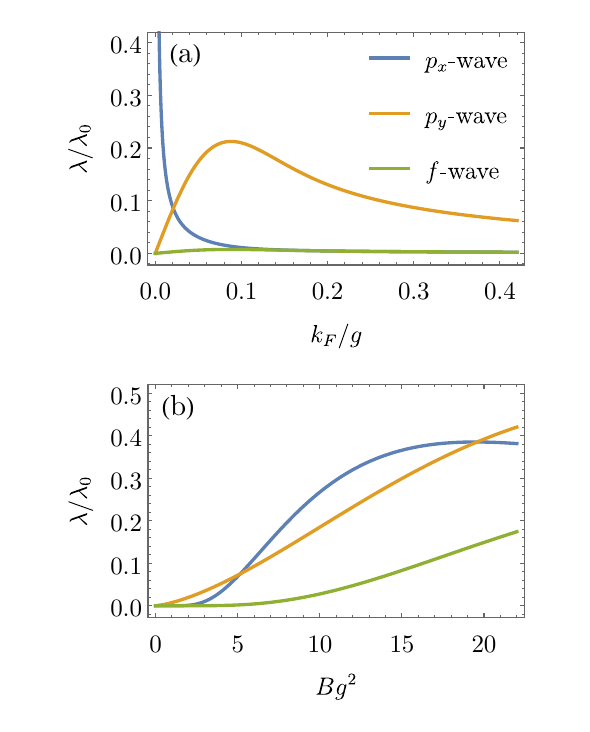}
    \caption{Plots of the eigenvalues (\ref{eq:lambda_px})--(\ref{eq:lambda_f}) of the LGE as a function of (a) $k_F$ with $Bg^2=5.5$ and (b) $B$ with $k_F/g=0.02$. A larger $\lambda$ means that the pairing channel is more stable. $\lambda_0=V_0gm$, $Ag^2=0.2$.}
    \label{fig:lambda}
\end{figure}

Under the small-$T_c$ (weakly interacting) assumption, the SC order parameter
\begin{equation}
    \Delta_{\mathbf{k}}=-\frac{1}{\mathcal{A}}\sum_{\mathbf{k}'}V_{\mathbf{k},\mathbf{k}'}\langle \hat c_{-\mathbf{k}'}\hat c_{\mathbf{k}'}\rangle,
\end{equation}
where $\hat c_{\mathbf{k}}$ is the annihilation operator in the lowest subband, can be obtained by solving the linearized gap equation restricted to the FS~\cite{Coleman_2015} (see SM \cite{SM} for a derivation).
Writing $\Delta_{(k_x,\pm k_F)}$ as $\Delta^{\pm}_{k_x}$, we have
\begin{multline}\label{eq:gapeq}
    \frac{1}{(2\pi)^2v_F}\left(\int_{-g/2}^{g/2} dk'_x V^{\pm+}_{k_x,k'_x} \Delta^+_{k'_x}+\int_{-g/2}^{g/2} dk'_x V^{\pm-}_{k_x,k'_x} \Delta^-_{k'_x}\right)\\=\lambda\Delta^\pm_{k_x},
\end{multline}
with the antisymmetry constraint $\Delta^-_{-k_x}=-\Delta^+_{k_x}$.
The candidate order parameters are (up to the highest harmonic appearing in $V$)
\begin{subequations}
    \begin{align}
    \label{eq:px-wave}
    \Delta^{\pm}_{k_x}&\propto\sin\frac{2\pi k_x}{g}& (p_x\text{-wave})\\
    \Delta^{\pm}_{k_x}&\propto\pm 1& (p_y\text{-wave})\\
    \Delta^{\pm}_{k_x}&\propto\pm\cos\frac{2\pi k_x}{g}& (f\text{-wave})\label{eq:f-wave}
\end{align}
\end{subequations}
with eigenvalues (larger means more stable in this convention).
\begin{subequations}\label{eq:lambda_s}
    \begin{align}
    \label{eq:lambda_px}\lambda_{p_x}&=\frac{V_0 g c_1}{(2\pi)^2v_F}c_2 \left[1+c_3(2k_F)\right],\\
    \lambda_{p_y}&=\frac{V_0 g c_1}{(2\pi)^2v_F}\left[1-c_3(2k_F)\right],\\
    \lambda_{f}&=\frac{V_0 g c_1}{(2\pi)^2v_F}c_2 \left[1-c_3(2k_F)\right].
    \label{eq:lambda_f}
\end{align}
\end{subequations}
Since $\lambda_{f}<\lambda_{p_y}$ is always true, we focus only on $p_x$ and $p_y$ in this Letter.
Unlike the isotropic system, the $p_x$ (sign changing along the transverse direction) and $p_y$ (sign changing along the stripe direction) pairings exhibit qualitatively distinct properties in our quasi-1D model. First, the $p_x$ pairing admits a nodal line at $k_x=0$, while the $p_y$ pairing is fully gapped with the gap function changing sign across the Fermi sheets. The longitudinal $p_y$ pairing can be viewed as arrays of 1D $p$-wave superconductors. Meanwhile, the transverse $p_x$ pairing is rarely discussed in the context of quasi-1D SC~\cite{Sengupta:2001} as the interwire interaction is required~\cite{Seroussi2014}. In our model, both pairings can be generated by the effective interaction given by Eq.~(\ref{eq:VN0}).

The eigenvalues of LGE [Eq.~(\ref{eq:lambda_s})] are crucially determined by $c_2$ and $c_3(2k_F)$ [given by Eq.~(\ref{eq:c_s})]. In the small $k_F$ limit with fixed $A$ and $B$, $c_3(2k_F)\to 1$, making $\lambda_{p_x}$ the largest and leading to the $p_x$-wave pairing. On the other hand, in the large $k_F$ limit, $c_3\to 0$, making $\lambda_{p_y}$ the largest (since $c_2<1$) and leading to the $p_y$-wave pairing. Therefore, a transition from $p_x$-wave pairing at small $k_F$ to $p_y$-wave at large $k_F$ is generically expected, as shown in in Fig.~\ref{fig:lambda}(a). The situation is more complicated when $k_F$ and $A$ are fixed. As shown in Figs.~\ref{fig:intro}(c) and \ref{fig:lambda}(a), the $p_x$ wave can become the leading pairing instability in the intermediate $B$ regime with a sufficiently small $k_F$. This nonmonotonicity arises from the $B$-dependence in $c_2$ and $c_3(2k_F)$, which reflects the anisotropic nature of the effective interaction described by Eq.~(\ref{eq:VN0}).

To obtain the phase boundary, we solve $\lambda_{p_x}=\lambda_{p_y}$, obtaining
\begin{equation}\label{eq:analytical-phase-boundary}
    k_F=\sqrt{\frac{A}{2AB+B^2}\ln\coth\left(\frac{\pi^2}{(2A+B)g^2}\right)}
\end{equation}
which is plotted as the boundary between two colors in Fig.~\ref{fig:intro}(c) (at this parameter regime, the error due to the approximations is invisible).
Although the asymptotics of Eq.~(\ref{eq:analytical-phase-boundary}) show that the preferred pairing would become $p_x$-wave at vanishingly small $Bg^2$ (e.g.\ would be $\sim10^{-16}$ in Fig.~\ref{fig:intro}), it is unphysical due to the failure of the $N=0$ approximation when $c_2$ becomes smaller than the Gaussian tails of the $N\neq0$ peaks.
At $B=0$ we have $V_{\mathbf{k},\mathbf{k}'}=V_{\mathbf{k},-\mathbf{k}'}$ without the $N=0$ approximation [this can be obtained from Eq.~(\ref{eq:VFinal}), but is more directly from Eq~(\ref{eq:Hintq}) as it vanishes when the form factors are unity], showing that there is no pairing.

The approximation above also neglects the imaginary part of the effective interaction due to the $N\neq0$ peaks, such that the real-valued symmetry $V_{q_x,q_y}=V_{q_x,-q_y}$ enforces the $p$-wave pairing to be either $p_x$- or $p_y$-wave. 
If we consider the imaginary parts, the symmetry becomes 
\begin{equation}
    V_{\mathbf{k},\mathbf{k}'}=V^*_{(k_x,-k_y),(k_x',-k_y')}
\end{equation}
and in general $p_x$- and $p_y$-wave mix.
One can fix a phase to have $\Delta_{\mathbf{k}}=\Delta^*_{(k_x,-k_y)}$, that is, the real and imaginary parts are $p_x$- and $p_y$-wave, respectively.
Hence, the transition in Eq.~(\ref{eq:analytical-phase-boundary}) is, in fact, a crossover that becomes very sharp when $Ag^2$ is small.
When $Ag^2$ is not small, the above approximation fails, and the crossover spreads out, with the small $k_F$ regime becoming $p+ip$ in the $A\to\infty$ (no stripe) limit \cite{May-MannJ2025,Li2025,Patri2025}.
See SM \cite{SM} for more discussion on this.

\textit{Discussion. ---} We study the possible single-flavor superconducting states in the presence of finite Berry curvature and a strong stripe potential, providing analytical solutions of two distinct pairings: longitudinal $p_y$ and transverse $p_x$ waves (where the stripes are along the $y$ direction), and constructing a phase diagram. The transverse $p_x$-wave SC is rare in quasi-1D SC~\cite{Sengupta:2001}, as it requires the inter-stripe pairing to dominate over the intra-stripe pairing. Our results establish a concrete, experimentally relevant way to realize such an unexpected state.

We now discuss how to distinguish the transverse $p_x$ and longitudinal $p_y$ waves experimentally. First, the transverse $p_x$ wave has a nodal line at $k_x=0$, while the longitudinal $p_y$ wave is fully gapped. Nodal superconductors generically permit low-energy quasiparticle excitations and lead to a reduction of superfluid stiffness at a finite current~\cite{Dahm1999,Yip1992,Xu1995}. As a result, measurements that probe the quasiparticle excitations (e.g., tunneling spectroscopy) can distinguish the two different orders (gapless for $p_x$ wave and fully gapped for $p_y$ wave). The electric transport can, in principle, provide evidence for the nodal SC \cite{Dahm1999}. The reduction of superfluid stiffness at a finite current tends to make the current depairing effect stronger for the transverse $p_x$ wave, and the nonlinear IV likely develops strong nonlinearity at very small applied current, contrary to the sharp threshold behavior for the fully gapped SC (e.g., the longitudinal $p_y$ wave). Notably, the applied currents here should be along the stripe in the transport characterization, as our theory focuses on the opened FS, leading to strong anisotropic transport signatures along two different directions (potentially relevant to~\cite{MorissetteE2025b}). Lastly, the phase sensitive measurement can provide decisive signatures of the symmetry of the gap by probing the phase along different angles.

\textit{Acknowledgement} --
This work is supported by the Laboratory for Physical Sciences.

\bibliographystyle{apsrev4-2}
   \bibliography{references}

	\newpage \clearpage 
	
	\onecolumngrid

    \setcounter{figure}{0}
	\renewcommand{\thefigure}{S\arabic{figure}}
	\setcounter{equation}{0}
	\renewcommand{\theequation}{S\arabic{equation}}

    \begin{center}
		{\large
			Stripe-tuned superconductivity in single-flavor metals with nontrivial quantum geometry
			\vspace{4pt}
			\\
			SUPPLEMENTAL MATERIAL
		}
	\end{center}
	
	In this supplemental material, we provide technical details for the main results presented in the main text.

\section{Detailed derivation of the sub-band strcture}

We start from the gauge transformed single-particle Hamiltonian (\ref{eq:H0transformed})
\begin{equation}
    H_0 = \sum_{\mathbf{k}}\frac{k^2}{2m}\tilde c^\dagger_\mathbf{k}\tilde c_\mathbf{k} - \tilde U_0\sum_\mathbf{k}\left(\tilde c^\dagger_{\mathbf{k}+(g,0)}\tilde c_\mathbf{k}+\tilde c^\dagger_{\mathbf{k}-(g,0)}\tilde c_\mathbf{k}\right).
\end{equation}
The single-particle Hamiltonian (in $\mathbf{k}$ basis) is block diagonal in $k_x$ (mod $g$) and $k_y$. Within a block, the Hamiltonian is
\begin{equation}
    H_{n,n'}=\frac{(k_x+ng)^2+k_y^2}{2m}\delta_{n,n'}-\tilde U_0(\delta_{n,n'-1}+\delta_{n,n'+1})
\end{equation}
which can be transformed to real space to obtain the following Mathieu-type 1D Schr\"odinger equation
\begin{equation}
    \left[\frac{1}{2m}\left(-\frac{d^2}{dx^2}+k_y^2\right)-2\tilde U_0\cos(gx)\right]\psi(x)=E\psi(x)
\end{equation}
with
\begin{equation}
    \psi(x)=e^{ik_x x}\sum_{n=-\infty}^\infty \psi_n e^{ingx}
\end{equation}
Comparing with the standard Mathieu characteristic equation with ``Floquet'' solutions
\begin{equation}
    y''+(a-2q\cos(2z))y=0,\quad y(z)=e^{irz}f(z)
\end{equation}
where $f(z)$ has a period $\pi$, we have
\begin{equation}
    a=\frac{8mE-4k_y^2}{g^2},\quad q=-\frac{8m\tilde U_0}{g^2},\quad z=\frac{g}{2}x,\quad r=\frac{2}{g}k_x
\end{equation}
The lowest subband corresponds to $r\in (-1,1)$ and the solution is $a=a_r(q)$ (the Mathieu characteristic value) and $y=\operatorname{ce}_r(q,z)+i\operatorname{se}_r(q,z)$ (Mathieu functions). Higher band energy and wavefunctions are the same expressions with $r$ being in $(-n,-n+1)\cup (n-1,n)$, corresponding to the $n$th excited state of the harmonic oscillator in the real-space tight binding limit.

Consider the limit $\frac{g^2}{2m}\ll\tilde U_0$ (that is, $2U_0m\gg g^2e^{\frac{B}{4}g^2}$), the lowest subband in the effective 1D model can be described by a tight-binding model where the electron is bound to the harmonic-oscillator-like wells near the local minima of the cosine potential.

This approximation is often used in the study of optical lattices~\cite{Arzamasovs2017}.
From the result in the reference, we have (note that $q<0$)
\begin{equation}
    a\approx a_{\frac{1}{2}}(q)-16\sqrt{\frac{2}{\pi}}(-q)^\frac{3}{4}e^{-4\sqrt{-q}}\cos(\pi r)
\end{equation}
\begin{equation}
    y(z)\approx e^{irz}\sum_{n=-\infty}^{\infty}e^{2inz}e^{-\frac{(2n+r)^2}{4\sqrt{-q}}}.
\end{equation}
Changing back to the physical variables, we arrive at the real-space tight-binding approximation of the system
\begin{equation}\label{eq:real-space-TB}
    E=E_0+\frac{k_y^2}{2m}-2t\left(\cos\frac{2\pi k_x}{g}-1\right)
\end{equation}
\begin{equation}\label{eq:hoppingStrength}
    t=\frac{1}{A^{3/2}gm}\sqrt{\frac{2}{\pi}}e^{-\frac{4}{Ag^2}}.
\end{equation}
\begin{equation}\label{eq:wavefunction}
    \hat\psi_{n,\mathbf{k}}=\left(\frac{2Ag^2}{\pi}\right)^{\frac{1}{4}}e^{-A(k_x+ng)^2-i\frac{B}{2}(k_x+ng)k_y}.
\end{equation}

Consider a chemical potential $\mu$ measured with respect to $E_0$.
At zero temperature, we then have the approximated FS
\begin{equation}
    k_y=\pm\sqrt{2m\mu+4mt\left(\cos\frac{2\pi k_x}{g}-1\right)}
\end{equation}
Letting $k_F=\sqrt{2m\mu}$ as in our convention, the range of variation of $k_y$ along the FS in the open-FS case is from $k_F$ to $\sqrt{k_F^2-8mt}$, and the critical FS between open and closed occurs at $k_F=\sqrt{8mt}$.

Since $t$ is exponentially small in $1/Ag^2$, the FS is open unless for exponentially small $k_F$.
Moreover, the variation of the open FS in the $x$ direction is also suppressed by $\sim\exp(-1/Ag^2)$, so we can normally ignore this variation and approximate the FS as two straight lines $k_y=\pm k_F$.
One subtlety here is that the effective interaction (\ref{eq:VN0}) also carries an exponential factor $\sim\exp\left(-\frac{1}{(A+\frac{B}{2})g^2}\right)$ controlling the $q_x$ dependence.
Nevertheless, unless $B\ll A$, the variation of FS is still much smaller than any feature scale of the interaction.
Thus, we will restrict our analytical treatment to a fixed $B>0$ (as well as a fixed $k_F$) in the $A\to0$ limit.

\section{Derivation of the effective interaction}

\begin{figure}
    \centering
    \includegraphics[trim={10 0 5 0},width=\linewidth]{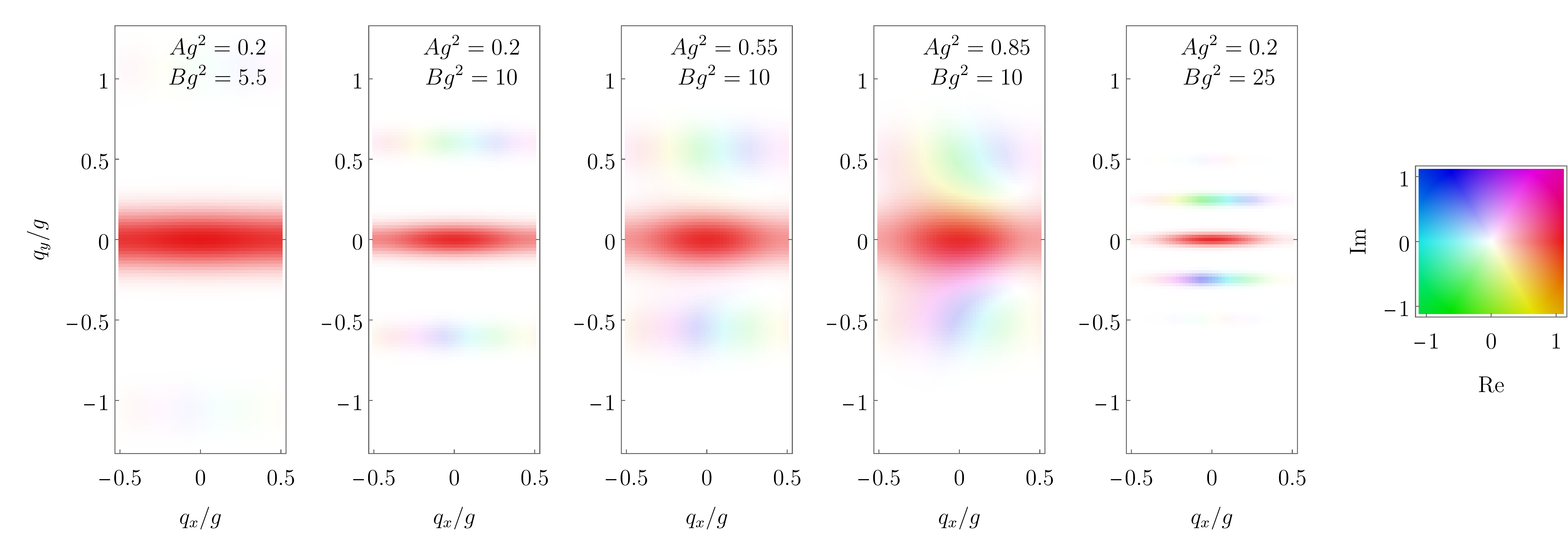}
    \caption{Plots of the effective interaction $V_{\mathbf{k}_1,\mathbf{k}_2,\mathbf{q}}$ in the lowest subband, with white and red corresponding to zero and positive, and hue being the complex phase. Here we choose $\mathbf{k}_1=-\mathbf{k}_2=(0.2,0.3)g$, but different choices only affect the phase shift in the ``rainbow'' parts. The plot is calculated numerically, but is almost visually indistinguisable from the analytical approximation (\ref{eq:Vanalytical}), except for $Ag^2=0.85$ where the $N=0$ and $N=\pm 1$ peaks are too close to each other that they start to deform from Gaussian. }
    \label{fig:interaction}
\end{figure}

Here we derive Eq.~(\ref{eq:VFinal}) in the main text.
We start from the interaction in the parent band
\begin{equation}\label{eq:Hint}
    H_\text{int}=-\frac{V_0}{2\mathcal{A}}\sum_{\mathbf{k}_1,\mathbf{k}_2,\mathbf{k}_3,\mathbf{k}_4}F(\mathbf{k}_1,\mathbf{k}_4)F(\mathbf{k}_2,\mathbf{k}_3)\delta_{\mathbf{k}_1+\mathbf{k}_2-\mathbf{k}_3-\mathbf{k}_4}c^\dagger_{\mathbf{k}_1}c^\dagger_{\mathbf{k}_2}c_{\mathbf{k}_3}c_{\mathbf{k}_4}
\end{equation}
To project it onto the lowest subband, we write the momenta in the form $\mathbf{k}+n\mathbf{g}$ with $\mathbf{k}$ in the first BZ, and replace $c^\dagger_{\mathbf{k}+n\mathbf{g}}\to\hat\psi^*_{n,\mathbf{k}}\hat{c}^\dagger_{\mathbf{k}}$. This gives
\begin{equation}\label{eq:projInt}
    \hat H_{\text{int}}=-\frac{1}{2\mathcal{A}}\sum_{\mathbf{k}_1,\mathbf{k}_2,\mathbf{k}_3,\mathbf{k}_4}V_{\mathbf{k}_1,\mathbf{k}_2,\mathbf{k}_3,\mathbf{k}_4}
    \,\bar\delta_{k_{1x}+k_{2x}-k_{3x}-k_{4x}}
    \delta_{k_{1y}+k_{2y}-k_{3y}-k_{4y}}\hat c^\dagger_{\mathbf{k}_1}\hat c^\dagger_{\mathbf{k}_2}\hat c_{\mathbf{k}_3}\hat c_{\mathbf{k}_4}
\end{equation}
where $\bar\delta_{k}:=\delta_{k\pmod g}$ and the effective interaction is
\begin{equation}\label{eq:Veff}
    V_{\mathbf{k}_1,\mathbf{k}_2,\mathbf{k}_3,\mathbf{k}_4}=V_0\sum_{n_1,n_2,n_3,n_4}
    F(\mathbf{k}_1+n_1\mathbf{g},\mathbf{k}_4+n_4\mathbf{g})F(\mathbf{k}_2+n_2\mathbf{g},\mathbf{k}_3+n_3\mathbf{g})
    \cdot\delta_{n_1+n_2-n_3-n_4} \hat\psi^*_{n_1,\mathbf{k}_1}\hat\psi^*_{n_2,\mathbf{k}_2}\hat\psi_{n_3,\mathbf{k}_3}\hat\psi_{n_4,\mathbf{k}_4}.
\end{equation}
We will restrict ourselves to the non-Umklapp case where $\mathbf{k}_1+\mathbf{k}_2-\mathbf{k}_3-\mathbf{k}_4=0$, which includes the Cooper pairing $\mathbf{k}_1+\mathbf{k}_2=\mathbf{k}_3+\mathbf{k}_4=0$ that we need.
With the change of variables
\begin{align}\label{eq:k1k2qtransf}
    \mathbf{k}_4&=\mathbf{k}_1+\mathbf{q},\quad n_4=n_1+m\\
    \mathbf{k}_3&=\mathbf{k}_2-\mathbf{q},\quad n_3=n_2-m,
\end{align}
and plug in the analytical wavefunctions (\ref{eq:wavefunction}), we have
\begin{equation}\begin{aligned}
    V_{\mathbf{k}_1,\mathbf{k}_2,\mathbf{q}}=V_0\frac{2Ag^2}{\pi}\sum_{n_1,n_2,m}\exp\Big\{
    &-\frac{B}{4}\left[(q_x+mg)^2+q_y^2\right]-i\frac{B}{2}\left[(k_{1x}+n_1g)q_y-k_{1y}(q_x+mg)\right]\\
    &-\frac{B}{4}\left[(q_x+mg)^2+q_y^2\right]+i\frac{B}{2}\left[(k_{2x}+n_2g)q_y-k_{2y}(q_x+mg)\right]\\
    &-A(k_{1x}+n_1 g)^2+i\frac{B}{2}(k_{1x}+n_1g)k_{1y}\\
    &-A(k_{2x}+n_2 g)^2+i\frac{B}{2}(k_{2x}+n_2g)k_{2y}\\
    &-A\left[k_{2x}-q_x+(n_2-m) g\right]^2-i\frac{B}{2}\left[k_{2x}-q_x+(n_2-m)g\right](k_{2y}-q_y)\\
    &-A\left[k_{1x}+q_x+(n_1+m) g\right]^2-i\frac{B}{2}\left[k_{1x}+q_x+(n_1+m)g\right](k_{1y}+q_y)\Big\}.
\end{aligned}\end{equation}
We can observe that the sum of $n_1$ (as well as $n_2$) is essentially an oscillating phase factor with a wide Gaussian envelope determined by $A$. Note that this should remain true even if we start with a different band structure. As long as we are in the real-space tight binding regime, we will have a wide envelope in momentum space, even if it is not Gaussian.

Now in this analytically tractable Gaussian case, we can single out these sums as Gaussian sums
\begin{equation}\begin{aligned}\label{eq:VGaussianSum}
    V_{\mathbf{k}_1,\mathbf{k}_2,\mathbf{q}}=V_0\frac{2Ag^2}{\pi}
    &\sum_m\exp\left\{-\frac{B}{2}\left[(q_x+mg)^2+q_y^2\right]-A(q_x+mg)^2-iB(k_{1x}-k_{2x}+q_x+mg)q_y\right\}\\
    &\sum_{n_1}\exp\left\{-iBgq_yn_1-2Ag^2\left(\frac{2k_{1x}+q_x+mg}{2g}+n_1\right)^2\right\}\\
    &\sum_{n_2}\exp\left\{iBgq_yn_2-2Ag^2\left(\frac{2k_{2x}-q_x-mg}{2g}+n_2\right)^2\right\}
\end{aligned}\end{equation}
so now it becomes clear that the speed of the oscillation part is determined by $q_y$. Since $Ag^2\ll 1$ in the real-space tight-binding regime, there are only small windows of $q_y$ that make all the sampled phases aligned to have a large final value; otherwise, the fast oscillation makes the sum essentially zero.
We use the Poisson summation formula to transform the second line of (\ref{eq:VGaussianSum}) into a sum over such peaks
\begin{equation}
    \sqrt{\frac{\pi}{2Ag^2}}\sum_{N_1}\exp\left\{i(Bgq_y+2\pi N_1)\frac{2k_{1x}+q_x+mg}{2g}-\frac{1}{8Ag^2}(Bgq_y+2\pi N_1)^2\right\}
\end{equation}
and similarly for the third line of (\ref{eq:VGaussianSum})
\begin{equation}
    \sqrt{\frac{\pi}{2Ag^2}}\sum_{N_2}\exp\left\{i(-Bgq_y+2\pi N_2)\frac{2k_{2x}-q_x-mg}{2g}-\frac{1}{8Ag^2}(-Bgq_y+2\pi N_2)^2\right\}
\end{equation}
Now when we combine the two sums, only the terms with $N_1=-N_2$ make significant contributions since $Ag^2\ll1$, and thus we can neglect other terms and turn it into a sum of $N=N_1=-N_2$.
After some algebra, we arrive at the final expression
\begin{equation}\label{eq:Vanalytical}
    V_{\mathbf{k}_1,\mathbf{k}_2,\mathbf{q}}=V_0\sum_m\exp\left\{-\left(A+\frac{B}{2}\right)(q_x+mg)^2\right\}
    \sum_N\exp\left\{i\frac{2\pi N}{g}(k_{1x}-k_{2x}+q_x)
    -\frac{1}{4Ag^2}(Bgq_y+2\pi N)^2-\frac{B}{2}q_y^2\right\}.
\end{equation}

Fig.~\ref{fig:interaction} shows the $V_{\mathbf{k}_1,\mathbf{k}_2,\mathbf{q}}$ calculated numerically from Eq.~(\ref{eq:Veff}), with the wavefunction $\hat{\psi}_{n,\mathbf{k}}$ directly calculated from Eq.~(\ref{eq:H0transformed}) without approximations.
Except for $Ag^2=0.85$, which is too large, other figures are visually indistinguishable from the analytical approximation (\ref{eq:Vanalytical}).
Note that the red part in the center is the $N=0$ peak, and the ``rainbow'' parts are the $N\neq0$ peaks.
For low $Bg^2=5.5$, all $N\neq0$ peaks are almost invisible; for $Bg^2=10$, only the $N=\pm 1$ rainbows are visible.
The $N=\pm2$ rainbows only start to be barely visible around $Bg^2=25$.

One can also intuitively see from this expression of the interaction that $p_y$ ($p_x$) wave pairing is preferred at large (small) $k_F$.
By approximating the FS as $k_y=\pm k_F$, the relevant $q_y$ in the scattering on the FS is either $0$ or $\pm 2k_F$ corresponding to two horizontal cuts in Fig.~\ref{fig:interaction}.
At large $k_F$, the variation of the interaction strengths within each cut is much smaller than the difference between the two cuts, making the interaction effectively a function of $q_y$, thus preferring $p_y$ pairing.
On the other hand, when $k_F$ is small, the two cuts become very close to each other, so that variation within the cut becomes more distinctive than the difference between the two cuts, making the interaction effectively a function of $q_x$ and preferring $p_x$ pairing.

To obtain Eq.~\eqref{eq:VN0} from Eq.~\eqref{eq:VFinal}, we first note that the sums over $m$ and $N$ factorize. Defining
\begin{equation}
    \alpha \equiv A+\frac{B}{2},
\end{equation}
the part that controls the dependence on $q_x$ is
\begin{equation}
    S_x(q_x)
    =
    \sum_{m\in\mathbb Z}
    e^{-\alpha(q_x+mg)^2}.
\end{equation}
Applying the Poisson summation formula,
\begin{equation}
    \sum_{m\in\mathbb Z}f(q_x+mg)
    =
    \frac{1}{g}
    \sum_{n\in\mathbb Z}
    e^{i2\pi n q_x/g}
    \widetilde f\left(\frac{2\pi n}{g}\right),
\end{equation}
with $f(x)=e^{-\alpha x^2}$ and
\begin{equation}
    \widetilde f(p)
    =
    \int_{-\infty}^{\infty}dx\,
    e^{-ipx}e^{-\alpha x^2}
    =
    \sqrt{\frac{\pi}{\alpha}}
    e^{-p^2/(4\alpha)},
\end{equation}
we obtain
\begin{align}
    S_x(q_x)
    &=
    \sqrt{\frac{\pi}{\alpha g^2}}
    \sum_{n\in\mathbb Z}
    e^{-\pi^2n^2/(\alpha g^2)}
    e^{i2\pi n q_x/g}
    \nonumber\\
    &=
    \sqrt{\frac{\pi}{\alpha g^2}}
    \left[
        1+
        2\sum_{n=1}^{\infty}
        e^{-\pi^2n^2/(\alpha g^2)}
        \cos\left(\frac{2\pi n q_x}{g}\right)
    \right].
    \label{eq:poisson-resummed-interaction}
\end{align}
Introducing
\begin{equation}
    c_1=
    \sqrt{\frac{\pi}{\left(A+\frac{B}{2}\right)g^2}},
    \qquad
    c_2=
    \exp\left[
        -\frac{\pi^2}
        {\left(A+\frac{B}{2}\right)g^2}
    \right],
\end{equation}
the exact harmonic expansion can be written compactly as
\begin{equation}
    S_x(q_x)
    =
    c_1
    \left[
        1+
        2\sum_{n=1}^{\infty}
        c_2^{\,n^2}
        \cos\left(\frac{2\pi n q_x}{g}\right)
    \right].
    \label{eq:harmonic-expansion}
\end{equation}
Keeping only the constant and first-harmonic terms, $n=0,\pm1$, gives
\begin{equation}
    S_x(q_x)
    \simeq
    c_1
    \left[
        1+
        2c_2\cos\left(\frac{2\pi q_x}{g}\right)
    \right].
\end{equation}

The second sum in Eq.~\eqref{eq:VFinal} determines the dependence on $q_y$ and on $k_{1x}-k_{2x}$. 
In the regime where the Gaussian envelope strongly suppresses the finite-$N$ peaks, we retain only the central peak $N=0$. 
The $N$-dependent phase factor then becomes unity, and the interaction no longer depends separately on $k_{1x}-k_{2x}$. 
The remaining factor is
\begin{align}
    S_y(q_y)
    &\simeq
    \exp\left[
        -\frac{(Bgq_y)^2}{4Ag^2}
        -\frac{B}{2}q_y^2
    \right]
    \nonumber\\
    &=
    \exp\left[
        -\left(
            \frac{B^2}{4A}+\frac{B}{2}
        \right)q_y^2
    \right]
    \equiv c_3(q_y).
\end{align}
Combining the first-harmonic approximation for $S_x$ with the central-peak
approximation for $S_y$ yields
\begin{equation}
    V_{\mathbf q}
    \simeq
    V_0c_1
    \left[
        1+
        2c_2\cos\left(\frac{2\pi q_x}{g}\right)
    \right]
    c_3(q_y),
\end{equation}
which is Eq.~\eqref{eq:VN0}.

The harmonic expansion in Eq.~\eqref{eq:harmonic-expansion} also has a
simple real-space interpretation. The real-space stripe period is
\begin{equation}
    a=\frac{2\pi}{g},
\end{equation}
so that the $n$th harmonic is $\cos(na q_x)$. 
Fourier transforming with respect to the momentum transfer $q_x$ shows that this term connects stripe orbitals whose centers are separated by $\pm na$. 
Thus, the $n$th harmonic generated by the Poisson summation corresponds to an interaction between stripes separated by $n$ real-space periods. In particular, the first harmonic describes nearest-neighbor-stripe interactions, whereas the second harmonic describes next-nearest-neighbor-stripe interactions. These are interaction or pair-scattering matrix elements rather than single-particle hopping amplitudes. Their coefficients decrease rapidly as $c_2^{\,n^2}$, which motivates truncating the interaction at the first harmonic. 

\section{The linearized gap equation}

In this appendix, we derive the linearized gap equation that we used to study pairing.
We begin by restricting the projected interaction (\ref{eq:projInt}) to the Cooper channel
\begin{equation}\label{eq:Hintcp}
    \hat{H}_\text{int,cp}=-\frac{1}{2\mathcal{A}}\sum_{\mathbf{k},\mathbf{k}'}V_{\mathbf{k},\mathbf{k}'}\hat c_{\mathbf{k}}^\dagger \hat c_{-\mathbf{k}}^\dagger \hat c_{-\mathbf{k}'}\hat c_{\mathbf{k}'}.
\end{equation}
where $\mathbf{k}=\mathbf{k}_1=-\mathbf{k}_2, \mathbf{k}'=-\mathbf{k}_3=\mathbf{k}_4$.
Define the SC order parameter
\begin{equation}
    \Delta_{\mathbf{k}}=-\frac{1}{\mathcal{A}}\sum_{\mathbf{k}'}V_{\mathbf{k},\mathbf{k}'}\langle \hat c_{-\mathbf{k}'}\hat c_{\mathbf{k}'}\rangle=-\frac{1}{\mathcal{A}}\sum_{\mathbf{k}'}\frac{V_{\mathbf{k},\mathbf{k}'}-V_{\mathbf{k},-\mathbf{k}'}}{2}\langle \hat c_{-\mathbf{k}'}\hat c_{\mathbf{k}'}\rangle.
\end{equation}
Note that we use the convention of summing over the entire $\mathbf{k}$ space rather than half of it, and we have $V_{-\mathbf{k},-\mathbf{k}'}=V_{\mathbf{k},\mathbf{k}'}$ and $\Delta_{-\mathbf{k}}=-\Delta_{\mathbf{k}}$.
Then, under the standard mean-field approximation, (\ref{eq:Hintcp}) becomes, up to a c-number,
\begin{equation}
    \hat{H}_\text{int}^\text{MF}=\frac{1}{2}\sum_{\mathbf{k}}\left(\Delta_{\mathbf{k}}\hat c_{\mathbf{k}}^\dagger \hat c_{-\mathbf{k}}^\dagger + \Delta^*_\mathbf{k}\hat c_{-\mathbf{k}}\hat c_\mathbf{k}\right)
\end{equation}
Now for each pair of $\mathbf{k},-\mathbf{k}$, the effective Hamiltonian for them at chemical potential $\mu$ measured with respect to $E_0$ is (again up to a constant)
\begin{equation}
    H_\mathbf{k}^\text{MF}=\begin{pmatrix}
        \hat c_\mathbf{k}^\dagger & \hat c_{-\mathbf{k}}
    \end{pmatrix}
    \begin{pmatrix}
        E_\mathbf{k}-\mu & \Delta_\mathbf{k}\\
        \Delta^*_\mathbf{k} & -E_\mathbf{k}+\mu
    \end{pmatrix}
    \begin{pmatrix}
        \hat c_\mathbf{k}\\
        \hat c^\dagger_{-\mathbf{k}}
    \end{pmatrix},
\end{equation}
from which we have, at temperature $T$ ($k_B=1$)
\begin{equation}
    \langle\hat c_{-\mathbf{k}}\hat c_\mathbf{k}\rangle=-\frac{\Delta_\mathbf{k}}{2\mathcal{E}_\mathbf{k}}\tanh{\left(\frac{\mathcal{E}_\mathbf{k}}{2T}\right)},\quad \mathcal{E}_\mathbf{k}=\sqrt{(E_\mathbf{k}-\mu)^2+|\Delta_\mathbf{k}|^2}
\end{equation}
So the self-consistency condition is
\begin{equation}
    \Delta_\mathbf{k}=\frac{1}{\mathcal{A}}\sum_{\mathbf{k}'}V_{\mathbf{k},\mathbf{k}'}\frac{\Delta_{\mathbf{k}'}}{2\mathcal{E}_{\mathbf{k}'}}\tanh{\left(\frac{\mathcal{E}_{\mathbf{k}'}}{2T}\right)} 
\end{equation}
Now when $T$ is only slightly below $T_c$ so that $|\Delta_\mathbf{k}|$ is small, we have $\mathcal{E}_\mathbf{k}\approx E_\mathbf{k}-\mu$.
So the final linearized gap equation (LGE) in the continuum is
\begin{equation}\label{eq:LGE1}
    \int\frac{d^2k'}{(2\pi)^2}\frac{1}{2(E_{\mathbf{k}'}-\mu)}\tanh\left(\frac{E_{\mathbf{k}'}-\mu}{2T_c}\right)V_{\mathbf{k},\mathbf{k}'}\Delta_{\mathbf{k}'}=\Delta_\mathbf{k}.
\end{equation}
Note that here we only look for $\Delta_\mathbf{k}$ up to an overall factor, which goes to zero as $T\to T_c$ from below.

When $T_c$ is small (which happens when $V_0$ is small), the LGE can be further simplified.
Write the integral in (\ref{eq:LGE1}) as
\begin{equation}
    I(T)=\int\frac{d^2k'}{(2\pi)^2}\frac{1}{2(E_{\mathbf{k}'}-\mu)}\tanh\left(\frac{E_{\mathbf{k}'}-\mu}{2T}\right)\frac{V_{\mathbf{k},\mathbf{k}'}}{V_0}\Delta_{\mathbf{k}'}.
\end{equation}
We will consider the $T\to0$ limit, which corresponds to taking $V_0\to0$ while keeping $T=T_c$ as a function of $V_0$.
Note that $E_\mathbf{k}$, $\mu$ and $V_{\mathbf{k},\mathbf{k}'}/V_0$ are fixed when taking the limit, so the only part that is changing is the $\tanh$ factor of the integrand.
Suppose the FS ($E_\mathbf{k}=\mu$) does not touch the VHS.
We choose a small ribbon $R$ of width $2r$ (independent of $T$) around the FS such that for $\mathbf{k}\in R$, we can approximate $E_\mathbf{k}\approx\mu+v_F(\mathbf{k}_\text{FS})k_\perp$, where $v_F$ denotes the Fermi velocity at a given momentum, $\mathbf{k}_\text{FS}$ is the projection of $\mathbf{k}$ onto the FS, and $k_\perp$ is the component of $\mathbf{k}$ perpendicular to the FS.
Moreover, this ribbon can be chosen to be small enough such that $V_{\mathbf{k},\mathbf{k}'}$ and $\Delta_{\mathbf{k}}$ do not vary too much in the direction perpendicular to the FS within the ribbon.
To approximate $I(T)$, we split into $I(T)=I_R(T)+I_{R^c}(T)$, where the two terms correspond to restricting the integral inside and outside $R$, respectively.
The first term can be approximated as
\begin{equation}
\begin{aligned}
    I_R(T)&\approx \frac{1}{(2\pi)^2}\int_\text{FS}dk_\text{FS}'\int_{-r}^rdk_\perp\frac{1}{2v_F(\mathbf{k}_\text{FS}')k_\perp}\tanh\left(\frac{v_F(\mathbf{k}_\text{FS}')k_\perp}{2T}\right)\frac{V_{\mathbf{k},{\mathbf{k}'_\text{FS}}}}{V_0}\Delta_{\mathbf{k}'_\text{FS}}\\
    &=\frac{1}{(2\pi)^2}\int_\text{FS}\frac{dk'}{v_F(\mathbf{k}')}\frac{V_{\mathbf{k},{\mathbf{k}'}}}{V_0}\Delta_{\mathbf{k}'}\int_0^{v_F(\mathbf{k}')r/2T}dx\, \frac{\tanh x}{x}\\
    &\sim \frac{1}{(2\pi)^2}\ln\left(\frac{1}{T}\right)\int_\text{FS}\frac{dk'}{v_F(\mathbf{k}')}\frac{V_{\mathbf{k},{\mathbf{k}'}}}{V_0}\Delta_{\mathbf{k}'}
\end{aligned}
\end{equation}
where in the last line we only keep the leading divergence in the $T\to 0$ limit.
On the other hand, the remaining part $I_{R^c}(T)$ in the $T\to0$ limit is,
\begin{equation}
    I_{R^c}(T)\to\int_{R^c}\frac{d^2k'}{(2\pi)^2}\frac{1}{2(E_{\mathbf{k}'}-\mu)}\operatorname{sign}\left(E_{\mathbf{k}'}-\mu\right)\frac{V_{\mathbf{k},\mathbf{k}'}}{V_0}\Delta_{\mathbf{k}'},
\end{equation}
which converges.
That is, the $T\to0$ limit of $I(T)$ is entirely governed by the log-divergence in $I_R(T)$ and does not depend on the size $r$ of the auxiliary ribbon.
The LGE then simplifies to
\begin{equation}\label{eq:gap-eq-FS1}
    \frac{1}{(2\pi)^2}\int_\text{FS}\frac{dk'}{v_F(\mathbf{k}')}V_{\mathbf{k},\mathbf{k}'}\Delta_{\mathbf{k}'}=\lambda\Delta_\mathbf{k}
\end{equation}
where $\lambda\sim\frac{1}{\ln(1/T_c)}$ (in practice, an additional energy scale, such as the bandwidth, is needed to determine $T_c$).
This is interpreted as an eigenvalue equation where the largest $\lambda$ with antisymmetric eigenvector corresponds to the most stable (largest $T_c$) pairing channel.
Equivalently, one can antisymmetrize $V_{\mathbf{k},\mathbf{k}'}$ and look for the largest eigenvalue.

\section{Derivation of the pairing channels}

In this section, we derive the pairing channels in the quasi-one-dimensional
limit and obtain Eqs.~\eqref{eq:px-wave}--\eqref{eq:lambda_f} of the main text.
Approximating the Fermi surface by two straight sheets at
$k_y=\pm k_F$, we define $\Delta_{k_x}^{\pm}
    \equiv
    \Delta_{(k_x,\pm k_F)}$.
The linearized gap equation then reduces to the two coupled integral equations
\begin{align}
    \lambda \Delta_{k_x}^{+}
    &=
    \mathcal C
    \int_{-g/2}^{g/2} dk_x'\,
    K(k_x,k_x')
    \left(
        \Delta_{k_x'}^{+}
        +c_3\Delta_{k_x'}^{-}
    \right),
    \label{eq:gap-plus}
    \\
    \lambda \Delta_{k_x}^{-}
    &=
    \mathcal C
    \int_{-g/2}^{g/2} dk_x'\,
    K(k_x,k_x')
    \left(
        c_3\Delta_{k_x'}^{+}
        +\Delta_{k_x'}^{-}
    \right),
    \label{eq:gap-minus}
\end{align}
where
\begin{equation}
    \mathcal C
    =
    \frac{V_0c_1}{(2\pi)^2v_F},
    \qquad
    K(k_x,k_x')
    =
    1+
    2c_2
    \cos\left[
        \frac{2\pi(k_x-k_x')}{g}
    \right].
\end{equation}

Because the kernel factorizes into a part acting on the two Fermi sheets and
a part acting on the longitudinal momentum, we seek solutions of the form
\begin{equation}
    \Delta_{k_x}^{s}
    =
    \eta_s\phi(k_x),
    \qquad
    s=\pm.
\end{equation}
The corresponding eigenvalue can be written as
\begin{equation}
    \lambda=a_\eta\kappa_\phi,
\end{equation}
where the sheet amplitudes satisfy
\begin{equation}
    \begin{pmatrix}
        1 & c_3 \\
        c_3 & 1
    \end{pmatrix}
    \begin{pmatrix}
        \eta_+ \\
        \eta_-
    \end{pmatrix}
    =
    a_\eta
    \begin{pmatrix}
        \eta_+ \\
        \eta_-
    \end{pmatrix}.
    \label{eq:sheet-eigenproblem}
\end{equation}
The two sheet eigenvectors and eigenvalues are
\begin{align}
    \eta_+=\eta_-:
    &\qquad
    a_+=1+c_3,
    \label{eq:sheet-symmetric}
    \\
    \eta_+=-\eta_-:
    &\qquad
    a_-=1-c_3.
    \label{eq:sheet-antisymmetric}
\end{align}
They correspond, respectively, to sheet-symmetric and sheet-antisymmetric
gap functions.

The longitudinal eigenvalue equation is
\begin{equation}
    \kappa_\phi\phi(k_x)
    =
    \mathcal C
    \int_{-g/2}^{g/2}dk_x'\,
    K(k_x,k_x')\phi(k_x').
    \label{eq:longitudinal-eigenproblem}
\end{equation}
Using
\begin{align}
    \cos\left[
        \frac{2\pi(k_x-k_x')}{g}
    \right]
    &=
    \cos\left(\frac{2\pi k_x}{g}\right)
    \cos\left(\frac{2\pi k_x'}{g}\right)
    \nonumber\\
    &\quad+
    \sin\left(\frac{2\pi k_x}{g}\right)
    \sin\left(\frac{2\pi k_x'}{g}\right),
\end{align}
we see that the kernel has support only in the three-dimensional harmonic
subspace
\begin{equation}
    1,
    \qquad
    \sin\left(\frac{2\pi k_x}{g}\right),
    \qquad
    \cos\left(\frac{2\pi k_x}{g}\right).
\end{equation}
These functions are mutually orthogonal over the interval
$[-g/2,g/2]$. In particular,
\begin{equation}
    \int_{-g/2}^{g/2}dk_x=g,
\end{equation}
and
\begin{equation}
    \int_{-g/2}^{g/2}dk_x\,
    \sin^2\left(\frac{2\pi k_x}{g}\right)
    =
    \int_{-g/2}^{g/2}dk_x\,
    \cos^2\left(\frac{2\pi k_x}{g}\right)
    =
    \frac{g}{2}.
\end{equation}
The longitudinal eigenfunctions and eigenvalues are therefore
\begin{align}
    \phi_0(k_x)&=1,
    &
    \kappa_0
    &=
    \frac{V_0gc_1}{(2\pi)^2v_F},
    \label{eq:kappa-constant}
    \\
    \phi_{\sin}(k_x)
    &=
    \sin\left(\frac{2\pi k_x}{g}\right),
    &
    \kappa_{\sin}
    &=
    \frac{V_0gc_1c_2}{(2\pi)^2v_F},
    \label{eq:kappa-sin}
    \\
    \phi_{\cos}(k_x)
    &=
    \cos\left(\frac{2\pi k_x}{g}\right),
    &
    \kappa_{\cos}
    &=
    \frac{V_0gc_1c_2}{(2\pi)^2v_F}.
    \label{eq:kappa-cos}
\end{align}
Here $\phi_0$ and $\phi_{\cos}$ are even under $k_x\rightarrow-k_x$,
whereas $\phi_{\sin}$ is odd.

For single-component fermions, the gap must satisfy
\begin{equation}
    \Delta_{-\mathbf k}=-\Delta_{\mathbf k}.
\end{equation}
Because inversion maps
\begin{equation}
    (k_x,+k_F)
    \longrightarrow
    (-k_x,-k_F),
\end{equation}
the two-sheet gap functions obey
\begin{equation}
    \Delta_{-k_x}^{-}
    =
    -\Delta_{k_x}^{+}.
    \label{eq:fermionic-constraint}
\end{equation}
For a factorized gap, this becomes
\begin{equation}
    \eta_-\phi(-k_x)
    =
    -\eta_+\phi(k_x).
    \label{eq:factorized-constraint}
\end{equation}
Consequently, an even longitudinal function must be combined with a
sheet-antisymmetric eigenvector, while an odd longitudinal function must be
combined with a sheet-symmetric eigenvector. This leaves three nonzero
pairing channels within the harmonic truncation of the interaction.

The first channel is uniform along each Fermi sheet but changes sign between
the two sheets:
\begin{equation}
    \Delta_{k_x}^{\pm}
    \propto
    \pm 1.
\end{equation}
This is the $p_y$-wave channel, with eigenvalue
\begin{equation}
    \lambda_{p_y}
    =
    a_-\kappa_0
    =
    \frac{V_0gc_1}{(2\pi)^2v_F}
    (1-c_3).
    \label{eq:lambda-py-derivation}
\end{equation}

The second channel is odd in $k_x$ and has the same sign structure on the two
Fermi sheets:
\begin{equation}
    \Delta_{k_x}^{\pm}
    \propto
    \sin\left(\frac{2\pi k_x}{g}\right).
\end{equation}
This is the $p_x$-wave channel, with eigenvalue
\begin{equation}
    \lambda_{p_x}
    =
    a_+\kappa_{\sin}
    =
    \frac{V_0gc_1c_2}{(2\pi)^2v_F}
    (1+c_3).
    \label{eq:lambda-px-derivation}
\end{equation}

Finally, the third channel changes sign between the two Fermi sheets and
also contains a nontrivial even harmonic along each sheet:
\begin{equation}
    \Delta_{k_x}^{\pm}
    \propto
    \pm \cos\left(\frac{2\pi k_x}{g}\right).
\end{equation}
It is therefore a higher-harmonic odd-parity channel, which we refer to as
the $f$-wave channel. Its eigenvalue is
\begin{equation}
    \lambda_f
    =
    a_-\kappa_{\cos}
    =
    \frac{V_0gc_1c_2}{(2\pi)^2v_F}
    (1-c_3).
    \label{eq:lambda-f-derivation}
\end{equation}
Unlike the fully gapped $p_y$ state, this channel has additional nodes at
$k_x=\pm g/4$, where
$\cos(2\pi k_x/g)=0$.

\section{Full phase diagrams in weak interacting approximation}

\begin{figure}
    \centering
    \includegraphics[width=0.9\linewidth]{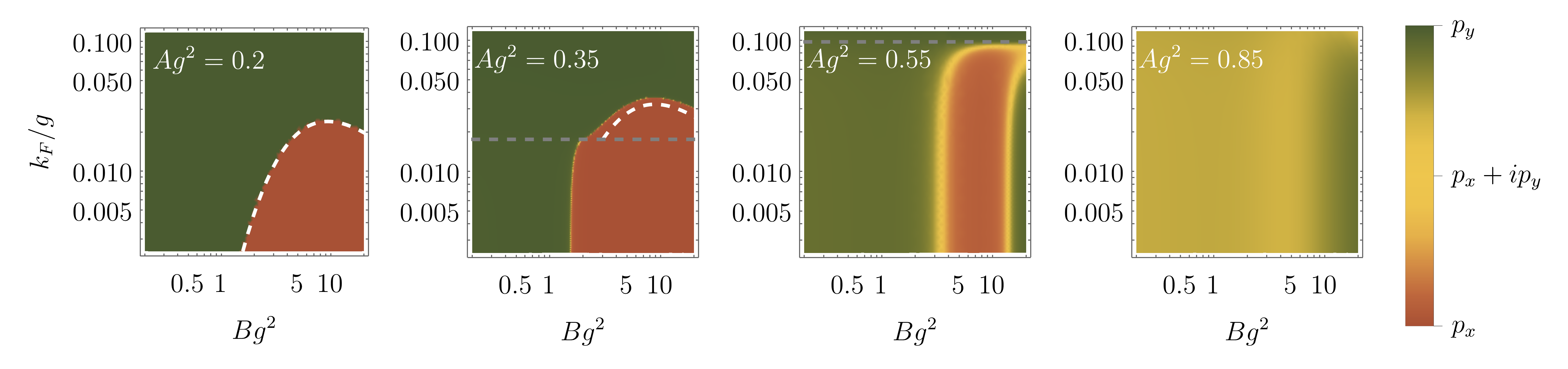}
    \caption{``Phase'' diagram of the preferred pairing form. White dashed lines indicate the analytical phase boundary (\ref{eq:analytical-phase-boundary}). Gray dashed line indicate the critical $k_F$ above (below) which the FS is open (closed).}
    \label{fig:phase}
\end{figure}

In this Appendix, we present the full ``phase'' diagram of the preferred pairing in the $(Ag^2,Bg^2,k_F/g)$ parameter space, by fixing $Ag^2$ and scanning through $(Bg^2,k_F/g)$.
For each parameter point, the effective interaction is directly calculated from Eq.~(\ref{eq:Veff}), with the wavefunction $\hat{\psi}_{n,\mathbf{k}}$ directly calculated from Eq.~(\ref{eq:H0transformed}) without approximation.
In this way, we can go to the regime where $Ag^2$ is not small, and thus the crossover behavior when the stripe disappears.
We numerically solve the FS version of LGE [Eq.~(\ref{eq:gap-eq-FS1})]
for the largest eigenvalue $\lambda$ with $\Delta_{-\mathbf{k}}=-\Delta_{\mathbf{k}}$.
Note that the FS can either be open or closed, depending on $k_F$ (recall that we measure $k_F$ along the positive $k_y$ axis).
By fixing the phase such that $\Delta_{k_x,k_y}=\Delta^*_{k_x,-k_y}$, we determine the pairing form parameter by
\begin{equation}
    \theta=\arg\left(\int_{\text{FS},k_x,k_y>0}\frac{dk}{v_F(\mathbf{k})}\Delta_\mathbf{k}\right)\mod \pi.
\end{equation}
That is, we determine the relative weight of the $p_x$-wave (real) and $p_y$-wave (imaginary) parts over the first quadrant of $k$-space.
In particular, $\theta=0,\frac{\pi}{4},\frac{\pi}{2},\frac{3\pi}{4}$ correspond to $p_x$, $p_x+ip_y$, $p_y$, $p_x-ip_y$ pairings, respectively.
Note that $f$-wave and higher harmonic components do not contribute to the value of $\theta$. In the regime we scan through, such higher-order components should be negligible, but not if we extend the diagrams to higher $B$ or larger $k_F$ (so that $Bk_F^2\gtrsim 1$).

The results are plotted in Fig.~\ref{fig:phase}.
For small $Ag^2=0.2$, the phase diagram agrees with the analytical results (\ref{eq:analytical-phase-boundary}) well.
For slightly larger $Ag^2=0.35$, the boundary between $\approx p_x$- and $\approx p_y$-wave regimes becomes fuzzy and deviates from the analytical boundary.
The crossover boundary in the closed FS (small $k_F$) regime is approximately vertical, which can be intuitively understood as the relative importance of $q_x$ and $q_y$ dependence being fixed when $V_\mathbf{q}$ is restricted to a small ellipse near $\mathbf{q}=0$.
For even larger $Ag^2=0.55$, all of the variation of the pairing form now happens only in the closed FS regime, with the boundary becoming fuzzier as $A$ grows larger (i.e., the stripe becomes weaker).
Finally, when $Ag^2\to\infty$, the entire $Bk_F^2\ll 1$ regime becomes $p_x+ip_y$ pairing as expected~\cite{May-MannJ2025}.

When $Bk_F^2\gtrsim 1$, the non-stripe case has higher angular momentum pairing~\cite{May-MannJ2025}.
With the stripe, this regime roughly corresponds to where the effect of the finite-$q_y$ ($N\neq 0$) peaks in $V_{\mathbf{k}_1,\mathbf{k}_2,\mathbf{q}}$ becomes essential, beyond just providing a small imaginary part that smooths out the transition between $p_x$- and $p_y$-wave pairing.
Such complicated rainbow-like peaks (see Fig.~\ref{fig:interaction}) lead to complicated complex pairing wave functions on the non-circular FS beyond $p$-wave.
This regime is not shown in Fig.~\ref{fig:phase}, but we believe that it has a direct crossover to the higher angular momentum regime without the stripe, similar to how the $p_x$-to-$p_y$ crossover smooths out to become $p_x+ip_y$ in the $Bk_F^2\ll 1$ regime.

\section{General interaction strengths and critical temperatures}

\begin{figure*}
    \centering
    \includegraphics[width=\linewidth]{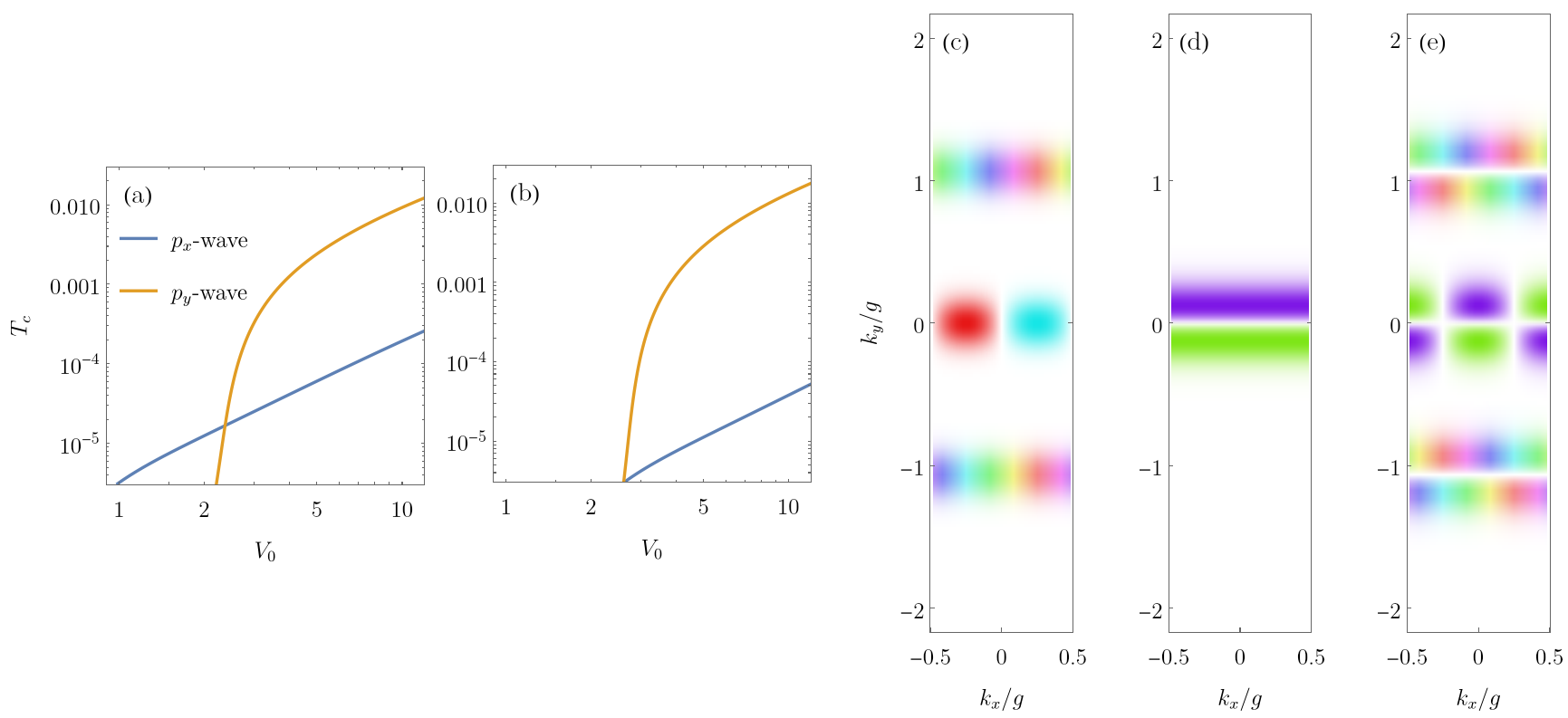}
    \caption{The SC $T_c$ (in units of $g^2/mk_B$) for different pairing waves as a function of interaction strength $V_0$ (in units of $1/m$) for (a) $Bg^2=5.5$, (b) $Bg^2=4$. (c) and (d) plots $\Delta_\mathbf{k}$ near the intersection of (a) for $p_x$-like and $p_y$-wave pairing, respectively, with colors scheme as in Fig.~\ref{fig:interaction} (note that $\Delta_{\mathbf{k}}$ is only defined up to a porportionality constant). (e) The $f$-wave-like paring corresponding to Eq.~(\ref{eq:f-wave}), near $V_0=66$, $T_c=1.8\times10^{-5}$ far away from dominated paring, shown here for completeness. Other parameters are $Ag^2=0.2$, $\mu=1.25\times 10^{-5}$ (corresponding to $k_F=0.005$ at $T=0$). The mixing between $p_x$- and $p_y$-wave near the FS is negligible in this parameter regime.  }
    \label{fig:full_kspace}
\end{figure*}

In this appendix, we go beyond the small-$V_0$ approximation and use the full 2D linearized gap equation to solve for the $T_c$ of the SC transition.
We will restrict ourselves to a small $Ag^2=0.2$ so that we can use the analytical form of the FS derived from Eqs.~(\ref{eq:real-space-TB}) and (\ref{eq:hoppingStrength}), and the analytical interaction form in Eq.~(\ref{eq:VFinal}).
Then we numerically solve the LGE without restricting to the FS [Eq.~(\ref{eq:LGE1})].
We numerically solve the full linearized gap equation without restricting
the momenta to the Fermi surface,
\begin{equation}
    \int \frac{d^2k'}{(2\pi)^2}
    \frac{1}{2(E_{\mathbf k'}-\mu)}
    \tanh\left(\frac{E_{\mathbf k'}-\mu}{2T}\right)
    \frac{V_{\mathbf k,\mathbf k'}}{V_0}
    \Delta_{\mathbf k'}
    =
    \chi(T)\Delta_{\mathbf k}.
    \label{eq:full-LGE-eigenvalue}
\end{equation}
Here the susceptibility $\chi(T)$ has units of inverse interaction strength,
$[\chi]=[V_0]^{-1}$. For a fixed interaction strength $V_0$, the
superconducting transition occurs when the largest eigenvalue satisfies
\begin{equation}
    V_0\chi_{\max}(T_c)=1.
\end{equation}

One numerical subtlety is that the $k_F$ corresponding to $\mu$ can be several orders of magnitude smaller than the scale at which $V_{\mathbf{k},\mathbf{k}'}$ begins to be suppressed by the Gaussian envelope.
This means that the $k_y$ integral has a long power-law tail until it being cut-off around $k_y\sim 1$.
In order to capture such features, we use a linear grid in $k_y$ at the scale of $k_F$ and a logarithmic grid from the scale of $k_F$ to that of $Bg^3$.

The result is shown in Fig.~\ref{fig:full_kspace}, with (a) being deeper in the $p_x$-wave regime of the $V_0\to0$ phase diagram than (b). We can see that at small $V_0$, $p_x$-wave pairing indeed has a larger $T_c$.
However, for larger $V_0$, $p_y$-wave has a larger $T_c$.
Also, when we move closer to the transition boundary in the $V_0\to0$ phase diagram from the $p_x$-wave side, the $V_0$ that prefers $p_x$-wave pairing will have a lower $T_c$, meaning that $p_x$-wave pairing becomes less and less stable.

\end{document}